\documentclass[prb,twocolumn,amsmath,amssymb,showpacs]{revtex4}

\usepackage{graphicx}% Include figure files
\usepackage{dcolumn}% Align table columns on decimal point
\usepackage{bm}% bold math
\usepackage{mathrsfs}
\usepackage{appendix}
\usepackage{bbm}
\usepackage{color}

\def \vS {{\bf S}}

\def \vQ {{\bf Q}}
\def \vq {{\bf q}}

\def \vm {{\bf m}}

\def \vH {{\bf H}}
\def \vF {{\bf F}}
\def \mb {\mu_{\rm B}}

\def \xp {{\bf x}^{\prime }}
\def \yp {{\bf y}^{\prime }}
\def \zp {{\bf z}^{\prime }}

\def \zz {{\bf z}}

\def \vP {{\bf P}}
\def \BF {{\rm BiFeO$_3$} }
\def \BP {{\rm BiFeO$_3$}}

\def \TN {T_{\rm N}}

\begin{document}

\title{Orientation Dependence of the Critical Magnetic Field for Multiferroic BiFeO$_3$}

\author{Randy S. Fishman}

\affiliation{Materials Science and Technology Division, Oak Ridge National Laboratory, Oak Ridge, Tennessee 37831, USA}

\date{\today}

\begin{abstract}

Multiferroic \BF undergoes a transition from a distorted spiral phase to a G-type antiferromagnet above a critical field $H_c$ that depends
on the orientation $\vm $ of the field.  We show that $H_c(\vm )$ has a maximum when oriented along 
a cubic diagonal parallel to the electric polarization $\vP $ and a minimum in the equatorial plane normal to $\vP $
when two magnetic domains with the highest critical fields are degenerate.  The measured critical field along a cubic axis is about 
19 T but $H_c$ is predicted to vary by as much as 2.5 T above and below this value.  The orientational dependence of  
$H_c(\vm )$ is more complex than indicated by earlier work, which did not consider the competition between magnetic domains.

\end{abstract}


\maketitle

Multiferroic materials offer the tantalizing prospect of controlling magnetic properties with an electric field and electric properties with a magnetic field.
Because of their technological promise, multiferroic materials remain the subject of intense interest.  Of all known multiferroic materials, only
\BF exhibits multiferroicity at room temperature.  As a ``proper" multiferroic, \BF has a ferroelectric transition temperature \cite{teague70}
$T_c \approx 1100$ K significantly higher than its N\'eel transition temperature \cite{sosnowska82} $\TN \approx 640$ K.  
Although the ferroelectric polarization is only slightly enhanced \cite{tokunaga10, park11} by the formation of a distorted spin cycloid, 
the magnetic domain distribution of \BF can be manipulated with an applied electric field  \cite{lebeugle08, slee08}.  

Using single crystals of \BF that have recently become available, inelastic neutron-scattering measurements \cite{jeong12, matsuda12, xu12} of the 
spin-wave spectra determined the nearest and next-nearest antiferromagnetic (AF) exchange interactions 
$J_1\approx 4.5$ meV and $J_2\approx 0.2$ meV on the psuedo-cubic unit cell with lattice constant \cite{moreau71} $a=3.96 \, \AA $.
Because the wavelength $\lambda \approx 62$ nm of the cycloid is so long \cite{sosnowska82, rama11a, herrero10, sosnowska11}, however,
inelastic neutron-scattering measurements \cite{matsuda12, fishman12} are unable to resolve the magnetic satellites at
wavevectors $(2\pi /a) (0.5\pm \delta , 0.5 \mp \delta ,0.5)$ ($\delta \approx 0.0045$), 
on either side of the antiferromagnetic (AF) Bragg wavevector $\vQ=(\pi /a )(1,1,1)$.  Consequently, inelastic neutron-scattering
cannot be used to determine the very small interaction energies of less than 1 meV that control the behavior of the cycloid.

By contrast, the spin-wave modes at the ordering wavevectors $\vQ_n$ of the cycloid can be measured very precisely with 
THz spectroscopy \cite{talbayev11, nagel13}.  The excellent agreement \cite{nagel13, fishman13b}
between the observed and predicted THz modes confirms that a microscopic model \cite{fishman13a}
with easy-axis anisotropy $K \approx 0.0035$ meV along the electric polarization direction 
$\zp =(1,1,1)$  (all unit vectors are assumed normalized to 1) and two Dzalyoshinskii-Moriya (DM) interactions can describe the multiferroic 
behavior of \BP .  Whereas the DM interaction $D \approx 0.107$ meV normal to the cycloidal plane fixes the cycloidal wavelength, 
the DM interaction \cite{kadomtseva04, ed05, pyatakov09, ohoyama11} $D'\approx 0.054$ meV along $\zp = (1,1,1)$ produces a 
small cycloidal tilt \cite{pyatakov09} that alternates in sign from one $[1,1,1]$ hexagonal plane to the next.

\begin{figure}
\includegraphics[width=8.5cm]{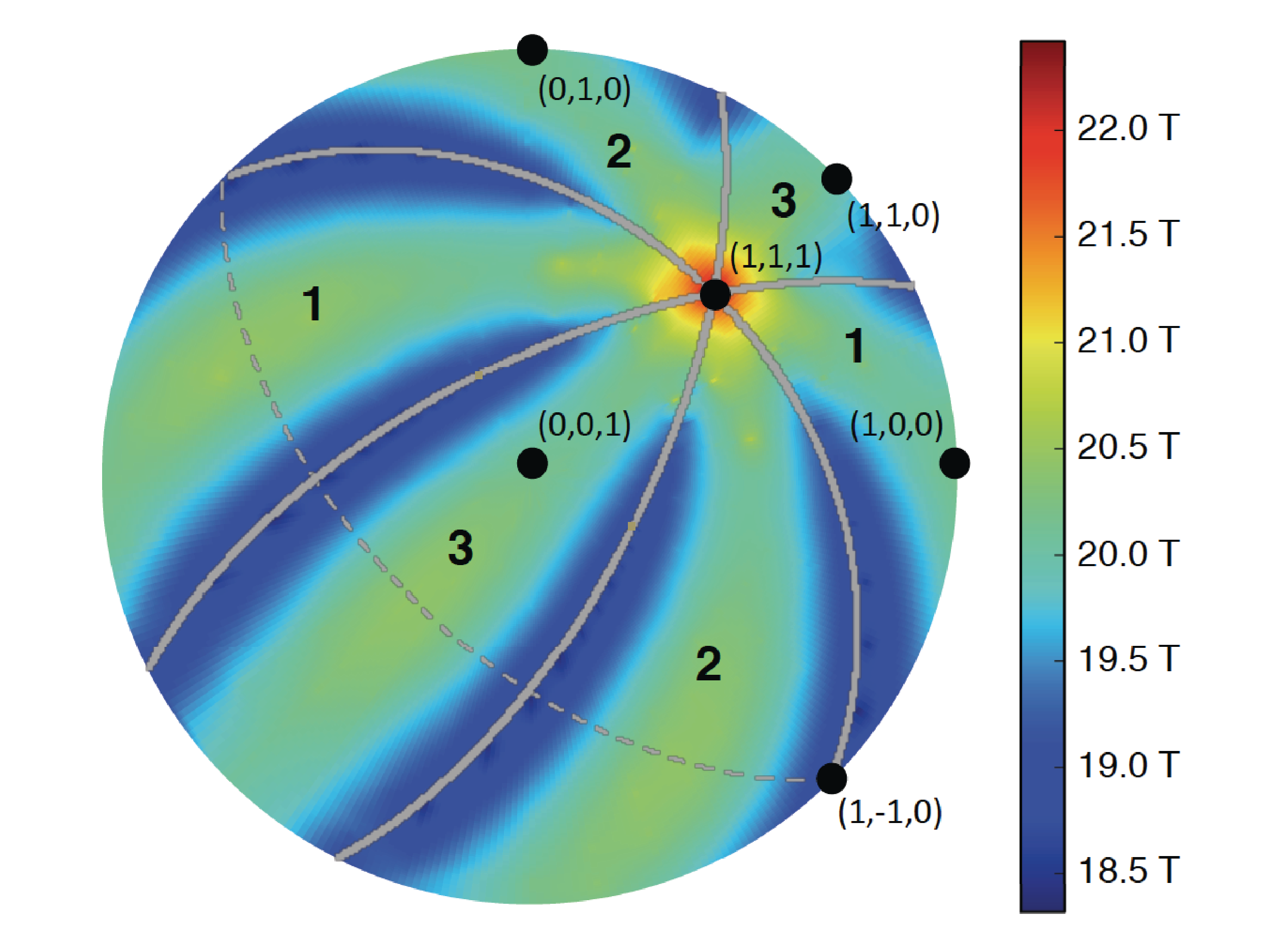}
\caption{The variation of $H_c$ over a hemisphere of the sphere for $\vm $.  $H_c$ is a maximum along the polarization direction $(1,1,1)$.
The tan lines are the borders between magnetic domains (denoted by 1, 2, or 3) with the highest $H_c$.  The dashed line is the equator with
the north pole given by $(1,1,1)$.
}
\end{figure}

If the DM and anisotropy interactions were absent, then $J_1$ and $J_2$ would stabilize a G-type AF with
ferromagnetic (FM) order within each $[1,1,1]$ hexagonal plane.  The distorted cycloid can be destabilized in favor of this 
antiferromagnet by chemical impurities \cite{chen12}, strain \cite{bai05}, and magnetic \cite{tokunaga10, park11} or electric fields \cite{sousa13}.
In the AF phase, the DM interaction $D'$ produces a weak FM moment \cite{tokunaga10, park11, rama11b} $M_0$ between 
0.03 and 0.06 $\mb $ per $S=5/2$ Fe$^{3+}$ ion due to the canting of the moments within each hexagonal plane. 

This paper uses the microscopic model described above to evaluate the critical magnetic field $H_c$ as a function of its 
orientation $\vm $.  In contrast to earlier theoretical work \cite{tokunaga10, bras09}, we find that the orientational 
dependence of $H_c(\vm )$ does not simply depend on the angles $\zeta $ and $\psi $ of the magnetic field with respect to the electric 
polarization.  While $H_c$ achieves a maximum when $\vm $ lies along $\zp $ ($\zeta \equiv \cos^{-1} (\vm \cdot \zp )=0$), it is a 
minimum in the equatorial plane normal to $\zp $ ($\zeta = \pi /2$) when the two magnetic domains 
with the highest critical fields are degenerate.  Because the qualitative predictions of this paper are unchanged by the precise 
interaction parameters, measurements of $H_c$ can be used to test the fundamental microscopic model described above.

With the electric polarization $\vP = P\zp $ along any of the eight equivalent cubic diagonals, the three magnetic domains of \BF have 
wavevectors $\vQ_n = \vQ + \vq_n$ where $\vq_1 = (2\pi /a)(0,\delta, -\delta )$, $\vq_2=(2\pi /a) (\delta ,0 -\delta )$, and 
$\vq_3=(2\pi/a)(\delta ,-\delta ,0)$.
For domain $n$, we construct a coordinate system with $\xp_n $ along $\vq_n$ and $\yp_n= \zp \times \xp_n $.
In zero field, the three domains are degenerate and equally occupied.  However, in a magnetic field, one of the domains has 
lower energy than the other two \cite{fishman13b}.  For $\vm =(1,0,0)$, domain 1 has the lowest energy and domains 2 and 3 
remain degenerate with higher energies.  Recent THz measurements \cite{nagel13} indicate that domains 2 and 3 are then depopulated 
above about 6 T.  Those measurements also indicate that it may be possible to reduce the population of the metastable domains 
by first applying a field far above $H_c$ and then reducing it to $H < 6$ T.

Generally, the domain with the lowest energy in a magnetic field $\vH = H \vm $ has the largest value of $\vert \yp_n \cdot \vm \vert $, so that the spins 
of that domain are predominantly perpendicular to the field (ignoring the small tilt produced by $D'$).  For a hemisphere of $\vm $ with
$\zp = (1,1,1)$, the solid curves in Fig.1 denote the boundaries between the domains with the lowest energies and the highest
critical fields.  Domains 1, 2, and 3 are degenerate when $\vm = \pm \zp $.

To obtain the critical field, we use the variational spin state described in earlier work \cite{fishman13b}.  
A cycloid with wavevector parameter $\delta =1/p$ has a wavelength of $\lambda = ap/\sqrt{2}$.
So with $p\gg 1 $ chosen to be an integer, the classical energy is minimized over a unit cell with two hexagonal layers, 
each containing $p$ sites.  A separate minimization loop is used to evaluate $p$ as a function of field.  
At zero field, $p=222$ gives a very good approximation to the measured value \cite{sosnowska82} $\delta \approx 0.0045$.  With increasing field, 
$\lambda \propto p$ grows.  Although they do not diverge at the first-order transition between the cycloidal and AF phases, 
$\lambda $ and $p$ increase by roughly a factor of three between $H=0$ and $H_c$.  

Beginning with the variational parameters known for zero field, $H$ is increased in increments of about 
0.015 T until the AF phase achieves a lower energy than the cycloidal phase, at which point the energies of both phases 
are interpolated to solve for $H_c(\vm )$.  This time-consuming procedure is required by the large number of variational parameters 
(including $p$) that determine the spin state.

Results for the critical field as a function of $\vm $ are given by the contours of Fig.1 over a hemisphere 
of $\vm $ with $(0,0,1)$ at the top of the hemisphere.  Notice that $H_c$ peaks at $\zp $ and reaches a minimum in the 
equatorial plane normal to $\zp $ at the borders between two degenerate domains.  We find that $H_c$ 
varies by about 4 T, from a minimum of 18.4 T to a maximum of 22.4 T.  Since $H_c(\vm )=H_c(-\vm)$, the results of 
Fig.1 can also be used to obtain $H_c(\vm )$ around $(0,0,-1)$ with another maximum at $-\zp $.

\begin{figure}
\includegraphics[width=8.5cm]{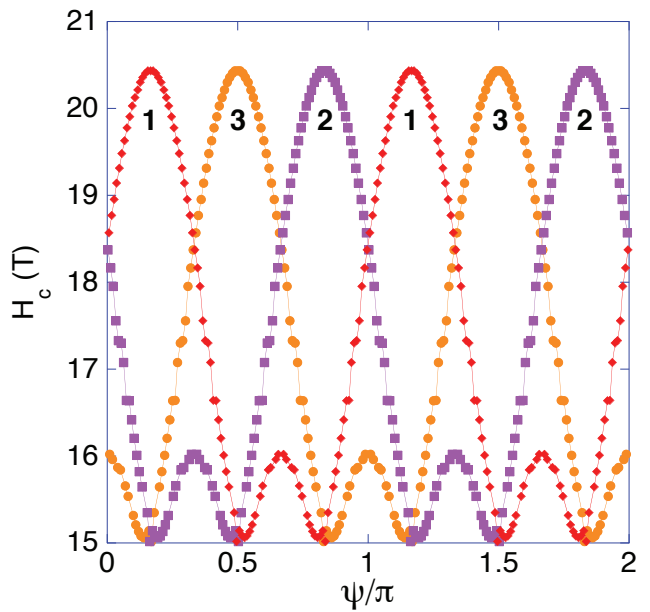}
\caption{The critical fields for domains 1 (circle), 2 (squares), and 3 (diamonds) versus $\psi $ in the equatorial 
$[1,1,1]$ plane normal to the polarization.}
\end{figure}

Previous models \cite{tokunaga10, bras09} found that $H_c(\vm )$ simply depends on the polar and azimuthal angles
$\zeta =\cos ^{-1}(\vm \cdot \zp )$ and $\psi = \cos^{-1}(\vm \cdot \xp_3)$.  
Tokunaga {\em et al.} \cite{tokunaga10} argued that $H_c(\vm )$ is a function only of $\zeta $ and is independent of $\psi $.   
Assuming a purely harmonic and coplanar cycloid, Le Bras {\em et al.} \cite{bras09} obtained a simple expression for the dependence of 
$H_c(\vm )$ on both $\zeta $ and $\psi $.  However, they did not consider the competition between cycloids in different magnetic domains.

The competition between magnetic domains produces the complex dependence of $H_c (\vm )$ 
on $\zeta $ and $\psi $.  Along the dashed equator ($\zeta = \pi /2$) sketched in Fig.1, the critical fields $H_c^{(n)}(\vm )$ for each 
domain are separately plotted versus $\psi $ in Fig.2.  While the individual critical fields 
$H_c^{(n)}(\vm )$ vary from 15.0 to 20.4 T, the maximum critical field $H_c(\vm )$ varies from 18.4 to 20.4 T.  In the equatorial plane, 
$H_c^{(n)}(\vm )$ is a maximum when $\vm = \pm \yp_n$, corresponding to azimuthal angles $\psi = -\pi/6 + n\pi/3 $ or $5\pi /6 +n\pi /3$.
When $H_c^{(n)}(\vm )$ reaches a maximum value, the critical fields for the other two domains reach their minimum values.
Because Le Bras {\em et al.} \cite{bras09} restricted consideration to a single magnetic domain, their predicted critical field
has a period of $\Delta \psi = \pi $ rather than $\Delta \psi = \pi /3$ as found here.

For $\vm $ along a cubic axis like $(0,0,1)$, several experimental groups \cite{tokunaga10, park11, nagel13} reported that
$H_c = 18.8$ T, which is about 1.4 T lower than our result.  To explain this disagreement, we examine the limitations
of our variational approach.  In equilibrium, the classical energy 
$E_i$ at each Fe$^{3+}$ site must be a minimum so that the forces $\vF_i = \partial E_i/\partial \vS_i$ on
the spin $\vS_i$ vanish.  The forces are quite small above $H_c$, indicating that the variational 
state provides an excellent description of the AF phase.  With increasing field below $H_c$, the forces
grow in magnitude as the variational spin state of the cycloid becomes compromised.  Because it provides an 
upper bound to the cycloidal energy, our variational approach will therefore underestimate rather than overestimate the critical field.
Hence, the limitations of this approach cannot explain the overestimation of the critical field.

But our classical model does not account for quantum spin fluctuations, which will differently affect the energies of the 
cycloidal and AF phases.  For the geometrically-frustrated antiferromagnet CuCrO$_2$, quantum fluctuations 
suppress the critical field \cite{fishman11} for the transition from a cycloidal to a collinear phase when the easy-axis anisotropy
is small.  In BiFeO$_3$, quantum fluctuations should also lower the critical field $H_c(\vm )$ from the classical values provided in this paper.

\begin{figure}
\includegraphics[width=9cm]{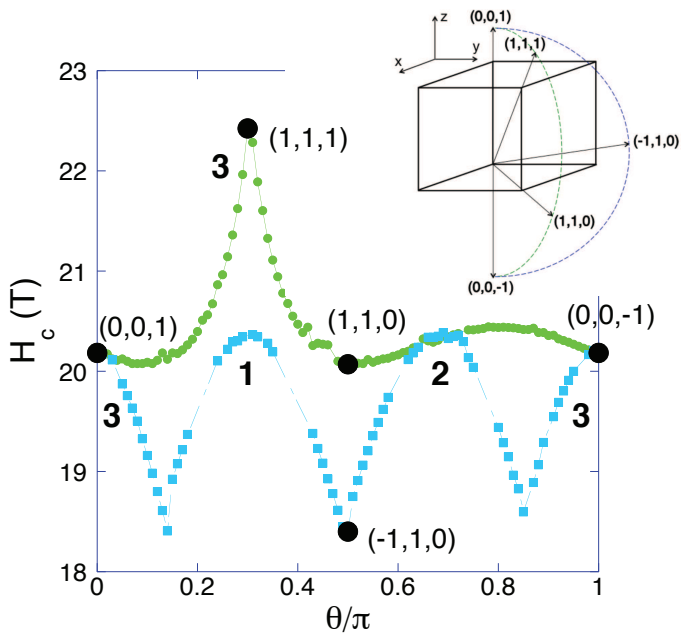}
\caption{The critical field for two longitudes connecting $(0,0,1)$ with (0,0,-1) through $(1,1,0)$ (circles) or through $(-1,1,0)$ (squares)
with $\theta = \cos^{-1} (\vm \cdot \zz )$.  The upper inset shows these two trajectories for the cubic unit cell.  }
\end{figure}

Comparing the predictions of this paper with measurements for $H_c(\vm )$ is complicated by several factors.  
Tokunaga {\em et al.} \cite{tokunaga10} observed hysterisis in the electric polarization due to the first-order transition from the 
cycloid to the AF.  Starting from above $H_c(\vm )$, Park {\em et al.} \cite{park11} reported that the jump in the 
electric polarization can occur at a slightly lower or higher field than the drop in the magnetization.  Whereas 
we have assumed that the critical field is produced by the domain $n$ with the lowest energy and highest $H_c^{(n)}(\vm )$,
experiments may detect the critical field for a different domain with higher energy.  These difficulties could 
explain some of the discrepancies between different experimental groups and between predictions and measurements.

For $\vm $ between $(0,0,1)$ and $(1,1,0)$, Tokunaga {\em et al.} \cite{tokunaga10} observed that $H_c(\vm )$ 
peaks at $\zp = (1,1,1)$ with a value of  24 T.  On the other hand, Fig.3 indicates that $H_c(\zp )=22.4$ T when 
$\theta =\cos^{-1}  (\zp \cdot \zz ) = 0.304\pi $ (54.7$^\circ $).   Notice that domain 3 has the lowest energy and highest 
critical field for all angles $\theta = \cos^{-1}  (\vm \cdot \zz ) $ along this longitude.  Tokunaga {\em et al.} also observed 
that $H_c(\vm ) = 18$ T for $\vm = (-1,1,0)$, a bit smaller than our prediction of 18.4 T.  Hence, their measurements yields a 
net range in $H_c$ of roughly 6 T, about 50\% larger than predicted in this paper.  By contrast, Park {\em et al.} \cite{park11} 
compared the critical fields for three different orientations $\vm $ and found that $H_c(\vm )= 19$ T is smallest when $\vm =\zp $.

More troubling, Tokunaga {\em et al.} \cite{tokunaga10} did not observe the predicted minimum in $H_c(\vm )$ 
between $(0,0,1)$ and $(-1,1,0)$ for $\theta = 0.166\pi $ (30$^\circ $), at the border between domains 3 and 1.  
This suggests that their sample may have been stuck within a fixed distribution of domains.  It may still be possible to 
observe the predicted minimum by applying and then removing a field far above $H_c(\vm )$ prior to each measurement.  

We have shown that, due to the competition between magnetic domains, the critical field $H_c(\vm )$ 
peaks at the polarization direction $\zp $ and passes through minima at the borders between domains.  Although measurements  
of $H_c(\vm )$ face several challenges, observation of the variation of $H_c(\vm )$
in the equatorial plane normal to $\zp $ would help to confirm the proposed microscopic model for \BP . 
We hope that this work will inspire more comprehensive measurements of the orientation
dependence of $H_c(\vm )$ for this highly important material.

I gratefully acknowledge conversations and technical assistance from Steven Hahn, Satoshi Okamoto,
and Toomas R\~o\~om.  Research sponsored by the U.S. Department of Energy, 
Office of Basic Energy Sciences, Materials Sciences and Engineering Division.

\end{document}